\documentclass[fleqn]{revtex4}

\usepackage{amsmath}
\usepackage{graphicx}

\begin{document}

\title{One-loop vacuum polarization at $m\alpha^7$ order for the two center problem}

\author{J.-Ph.~Karr}
\author{L.~Hilico}
\affiliation{Laboratoire Kastler Brossel, UPMC-Univ. Paris 6, ENS, CNRS, Coll\`ege de France\\
4 place Jussieu, F-75005 Paris, France}
\affiliation{Universit\'e d'Evry-Val d'Essonne, Boulevard Fran\c cois Mitterrand, F-91000 Evry, France}
\author{Vladimir I. Korobov}
\affiliation{Bogoliubov Laboratory of Theoretical Physics, Joint Institute
for Nuclear Research, Dubna 141980, Russia}

\begin{abstract}
We present calculations of the one-loop vacuum polarization contribution (Uehling potential) for the two-center problem in the NRQED formalism. The cases of hydrogen molecular ions ($Z_1=Z_2=1$) as well as antiprotonic helium ($Z_1=2$, $Z_2=-1$) are considered. Numerical results of the vacuum polarization contribution at $m\alpha^7$ order for the fundamental transitions $(v=0,L=0)\to(v'=1,L'=0)$ in H$_2^+$ and HD$^+$ are presented.
\end{abstract}

\maketitle

\section*{Introduction}

In Refs.~\cite{Korobov14PRA,Korobov14} a complete set of $m\alpha^7$-order contributions has been evaluated for the fundamental transitions of the hydrogen molecular ions H$_2^+$ and HD$^+$ as well as for two-photon transitions of antiprotonic helium. All calculations at this order were performed in the nonrecoil limit, by evaluating the one-electron QED corrections in the two-center approximation. The only exception is the Uehling potential vacuum polarization contribution, which was computed with a lower level of accuracy. Following the notations of Ref.~\cite{CODATA10}, Eq.(46), the Uehling correction at $m\alpha^7$ order for a two-center system can be written as
\begin{equation}\label{VP}
\Delta E_{vp}^{(7)} =
\frac{\alpha^5}{\pi}
   \biggl[ V_{61}\ln(Z\alpha)^{-2} + G_{VP}^{(1)}(R) \biggr]
   \left\langle V_{\delta} \right\rangle,
\end{equation}
where $R$ is the internuclear distance, and
\begin{equation}\label{distrib}
V_{\delta} (\mathbf{r}) =
   \pi \left[ Z_1^3 \delta(\mathbf{r_1}) + Z_2^3 \delta(\mathbf{r_2}) \right].
\end{equation}
The $V_{61}$ coefficient is known analytically, while the nonlogarithmic term was calculated in~\cite{Korobov14PRA,Korobov14} in the Linear Combination of Atomic Orbitals (LCAO) approximation using the hydrogen atom ground state value of $G_{VP}^{(1)}$. In this work we present a complete account of the vacuum polarization contribution in the two Coulomb center approximation.

We utilize the formalism of nonrelativistic quantum electrodynamics (NRQED), a similar approach has been used in \cite{PacPRC11} (see Sec.~II.B of that paper) for pionic hydrogen. We start from the nonrelativistic wave function and then obtain contributions due to the relativistic corrections to the electron wave function and modification of the Coulomb vertex function. This approach is first illustrated by calculating the Uehling potential energy shift for $S$-states of the hydrogen atom in Sec.~I.

Sec.~II extends the formalism to the two-center case, and the $G_{VP}^{(1)}(R)$ function is calculated. More precisely, the calculated terms also include all contributions of higher order in $\alpha$ generated by the Uehling potential and leading relativistic corrections. Final results for the fundamental transitions in the H$_2^+$ and HD$^+$ ions are presented and discussed in Sec.~III.

We use atomic units throughout.

\section{Hydrogen atom} \label{hydrogen}

In the NRQED formalism, the zero-order approximation is the nonrelativistic (Schr\"odinger) wave function $\Psi_0$ with Pauli spinors, defined by
\begin{equation}\label{NR}
\left(H_0-E_0\right)\Psi_0 = 0,
\qquad
H_0 = \frac{\mathbf{p}^2}{2}+V,
\qquad
V = -\frac{Z}{r}.
\end{equation}
For higher-order terms the Rayleigh-Schr\"odinger perturbation theory is used. If one wants to evaluate the one-loop vacuum polarization contribution to the bound electron in the external Coulomb field to the required $m\alpha^7$ order, one needs to evaluate the first-order contribution, which is the Uehling potential $U_{vp}(r)$ (Fig.~1a). Next is the leading-order relativistic correction to the wave function of the electron (Fig.~1b), which produces a second-order contribution with the Breit-Pauli Hamiltonian:
\begin{equation}\label{H_BP}
H_B = -\frac{p^4}{8}+\frac{1}{8}\Delta V,
\end{equation}
as the perturbation. The last term is the vertex function modification (Fig.~1c). The only contribution at this order to the vertex with the Coulomb photon interaction is the Darwin term, see Fig.~3 in \cite{Kinoshita96} or Eq.~(7) of \cite{Korobov09}.

In atomic units the Uehling potential is expressed:
\begin{equation}
U_{vp}(r) = -\frac{2}{3}\frac{Z\alpha}{\pi r}
   \int_1^\infty dt\> e^{-\frac{2r}{\alpha}\,t}
   \left(
      \frac{1}{t^2}+\frac{1}{2t^4}
   \right)
   \left(t^2-1\right)^{1/2}.
\end{equation}

Evaluation of the first-order correction with the nonrelativistic wave functions of the hydrogen $S$-states is straightforward and results in the following expression:
\begin{equation}\label{1st_order}
\Delta E_{vp}^{a} =
\left\langle
   nl\bigl|U_{vp}\bigr|nl
\right\rangle =
   \frac{\alpha(Z\alpha)^4}{\pi n^3}\left[
      -\frac{4}{15}
      +\frac{5\pi}{48}(Z\alpha)
      -\frac{2}{7}\left(1+\frac{1}{5n^2}\right)(Z\alpha)^2
      +\frac{\pi}{768}\left(49+\frac{35}{n^2}\right)(Z\alpha)^3+\dots
   \right]
\end{equation}

\begin{figure}
\begin{center}
\begin{tabular}{c@{\hspace{15mm}}c@{\hspace{15mm}}c}
$eU_{vp}(\mathbf{k}^2)$ & $~~~H_B$\hspace{9mm}$eU_{vp}(\mathbf{k}^2)$ &
$e(\mathbf{k}^2/8m_e)U_{vp}(\mathbf{k}^2)$\\[2mm]
\includegraphics[width=0.20\textwidth]{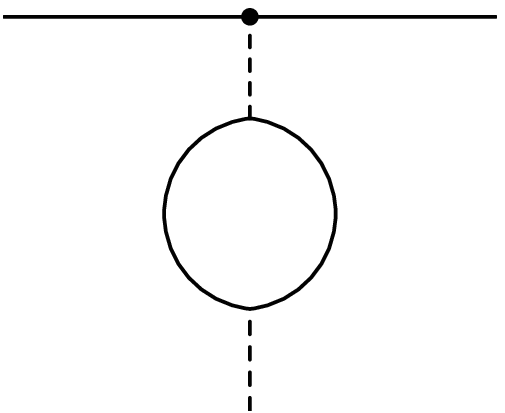}
&
\includegraphics[width=0.215\textwidth]{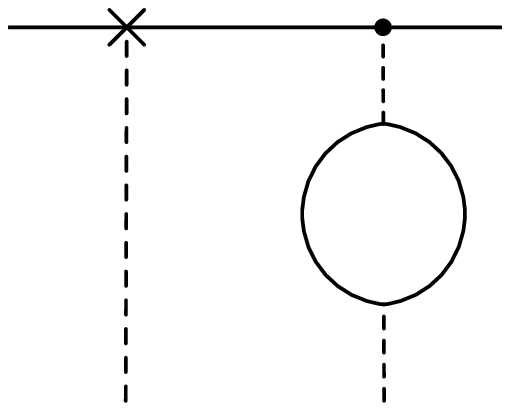}
&
\includegraphics[width=0.20\textwidth]{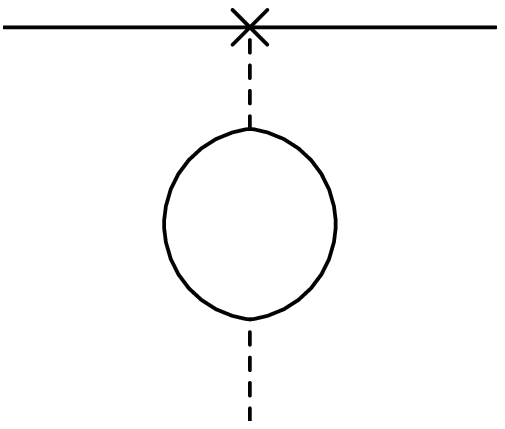}\\[2mm]
a) & b) & c)
\end{tabular}
\end{center}
\caption{Feynman diagrams for the one-loop vacuum polarization NRQED contributions.}
\end{figure}

The second-order term, determined by the diagram in Fig.~1.b, has a form
\begin{equation}\label{2nd_order}
\Delta E_{vp}^{b} =
2\left\langle
   \Bigl(H_B-\left\langle H_B \right\rangle\Bigr)(E_0-H)^{-1}
   \Bigl(U_{vp} - \left\langle U_{vp} \right\rangle\Bigr)
\right\rangle
\end{equation}
and may be obtained by substituting
$\Psi_B=(E_0-H)^{-1}\left(H_B-\left\langle H_B \right\rangle\right)\Psi_0$ into Eq.~(\ref{2nd_order}). An analytical expression of $\Psi_B$ can be found e.g. in~\cite{Korobov09}. For the $S$ states, one gets
\begin{equation}\label{2nd_order_S}
\begin{array}{@{}l}\displaystyle
\Delta E_{vp}^{b} =
   \frac{\alpha(Z\alpha)^4}{\pi n^3}
   \biggl\{
      -\frac{3\pi}{16}(Z\alpha)
\\[3mm]\displaystyle\hspace{15mm}
      -\frac{2}{15}\left[
         \ln{(Z\alpha)^{-2}}
         -2\left(
            \psi(n\!+\!1)-\psi(1)
            -\ln{n}+\ln{2}
            -\frac{107}{60}-\frac{2}{n}+\frac{5}{2n^2}
         \right)
      \right](Z\alpha)^2
\\[3mm]\displaystyle\hspace{15mm}
      +\frac{5\pi}{96}\left[
         \ln{(Z\alpha)^{-2}}
         -2\left(
            \psi(n\!+\!1)-\psi(1)
            -\ln{n}-\ln{2}
            -\frac{43}{60}-\frac{2}{n}+\frac{3}{n^2}
         \right)
      \right](Z\alpha)^3
      +\dots
   \biggr\}
\end{array}
\end{equation}
where $\psi$ is the logarithmic derivative of the Euler gamma function $\Gamma(z)$.

As discussed above, the NRQED effective Hamiltonian at $m\alpha(Z\alpha)^6$ order contains just one contribution determined by the diagram in Fig.~1c:
\begin{equation}\label{eff_H}
H^{(7)}_{vp} = \frac{1}{8}\Delta U_{vp}.
\end{equation}
Using
\[
\Delta\left(\frac{e^{-\Lambda r}}{r}\right) =
   -4\pi\delta(\mathbf{r})
   +\Lambda^2\>\frac{e^{-\Lambda r}}{r},
\]
one gets
\begin{equation}\label{Darwin}
H^{(7)}_{vp}
= -\frac{1}{12}\frac{Z\alpha}{\pi}
   \int_1^\infty dt\>
   \left[-4\pi\delta(\mathbf{r})+\frac{4t^2}{\alpha^2}\frac{e^{-\frac{2r}{\alpha}\,t}}{r}\right]
   \left(
      \frac{1}{t^2}+\frac{1}{2t^4}
   \right)
   \left(t^2-1\right)^{1/2}.
\end{equation}

Taking the expectation values of this effective Hamiltonian, one immediately gets for $S$ states
\begin{equation}
\Delta E_{vp}^c = \frac{1}{8}
\left\langle
   nl\bigl|\bigl(\Delta U_{vp}\bigr)\bigr|nl
\right\rangle =
   \frac{\alpha(Z\alpha)^4}{\pi n^3}
   \left[
      \frac{3\pi}{16}(Z\alpha)
      -\frac{1}{3}\left(1+\frac{1}{5n^2}\right)(Z\alpha)^2
      +\frac{5\pi}{576}\left(7\!+\!\frac{5}{n^2}\right)(Z\alpha)^3+\dots
   \right]
\end{equation}

The NRQED contribution, which is determined by the three terms of Fig.~1, should be exact up to $m\alpha(Z\alpha)^7$ order. The sum of these three contributions for $S$ states gives the final result
\begin{equation}\label{Uehling_fin}
\begin{array}{@{}l}\displaystyle
\Delta E_U =
   \frac{\alpha(Z\alpha)^4}{\pi n^3}
   \biggl\{
      -\frac{4}{15}
      +\frac{5\pi}{48}(Z\alpha)
      -\frac{2}{15}(Z\alpha)^2\ln{(Z\alpha)^{-2}}
\\[3mm]\displaystyle\hspace{15mm}
      +\frac{4}{15}(Z\alpha)^2
      \left[
         \psi(n\!+\!1)-\psi(1)
         -\ln{\left(\frac{n}{2}\right)}
         -\frac{431}{105}-\frac{2}{n}+\frac{57}{28n^2}
      \right]+\dots
\\[3mm]\displaystyle\hspace{15mm}
      -\frac{2}{15}\left[
         \ln{(Z\alpha)^{-2}}
         -2\left(
            \psi(n\!+\!1)-\psi(1)
            -\ln{n}+\ln{2}
            -\frac{431}{105}-\frac{2}{n}+\frac{57}{28n^2}
         \right)
      \right](Z\alpha)^2
\\[3mm]\displaystyle\hspace{15mm}
      +\frac{5\pi}{96}\left[
         \ln{(Z\alpha)^{-2}}
         -2\left(
            \psi(n\!+\!1)-\psi(1)
            -\ln{n}-\ln{2}
            -\frac{153}{80}-\frac{2}{n}+\frac{103}{48n^2}
         \right)
      \right](Z\alpha)^3
      +\dots
   \biggr\}
\end{array}
\end{equation}
The first three lines are in complete agreement with the combined result of~\cite{Mohr_VP,Karsh97}. The last line extends the general expression of $\Delta E_U$ by one further order in $Z\alpha$.

\section{Two-center problem}
Now, we are ready to study two-center systems. The nonrelativistic Hamiltonian of an electron is then
\begin{equation}
H_0 = \frac{p^2}{2} + V, \hspace{3mm} V = -\frac{Z_1}{r_1} - \frac{Z_2}{r_2}.
\end{equation}
The energy and wavefunction of the ground ($1s\sigma$) state will be denoted by $E_0$ and $\psi_0$ respectively. The Uehling potential is a sum of interactions with both nuclei:
\begin{equation}
U_{vp}(\mathbf{r}) = U_{vp}(r_1) + U_{vp}(r_2),
\end{equation}
%
%
%
We now want to calculate the contributions corresponding to diagrams a), b) c) of Fig.~1 in the same way as it was done in the previous Section for the hydrogen atom.

The first of these diagrams contains the leading-order contributions (of orders $\alpha(Z\alpha)^4$ and $\alpha(Z\alpha)^5$) which were already included in earlier calculations~\cite{Korobov06}. We are thus interested in higher-order ($\alpha(Z\alpha)^6$ and above) terms, which can be obtained by the following subtraction:

\begin{eqnarray}
\Delta E_a^{(7+)} &=&
   \left\langle \psi_0 | U_{vp} | \psi_0 \right\rangle - \Delta E_{vp}^{(5)} - \Delta E_{vp}^{(6)}  \\
&=& \left\langle \psi_0 | U_{vp} | \psi_0 \right\rangle
   + \frac{4\alpha^3}{15} \left\langle Z_1 \delta(\mathbf{r_1}) + Z_2 \delta(\mathbf{r_2}) \right\rangle
   - \frac{5\alpha^4}{48} \pi \left\langle Z_1^2 \delta(\mathbf{r_1}) + Z_2^2 \delta(\mathbf{r_2}) \right\rangle
\label{E7a} \nonumber
\end{eqnarray}
As shown in Sec.~\ref{hydrogen}, diagrams b) and c) both contain $\alpha(Z\alpha)^5$-order terms, which cancel each other. Writing $\Delta E_b$ in terms of the first-order perturbation wavefunction $\psi_B$ associated with the Breit-Pauli Hamiltonian
\begin{equation}
\Delta E_b = 2 \left\langle \psi_B | U_{vp} | \psi_0 \right\rangle
\end{equation}
with
\begin{equation}
(E_0 - H_0 ) \psi_B = \left( H_B - \langle H_B \rangle \right) \psi_0 \; ,
\end{equation}
one can see that the $\alpha(Z\alpha)^5$-order term in $\Delta E_{b}$ comes from the leading $1/r$ singularity of $\psi_B$. In order to get the contribution of order $\alpha(Z\alpha)^6$ and above, it is convenient to subtract this singularity and use the wavefunction $\tilde{\psi}_B$ defined by
\begin{equation}
\psi_B = \tilde{\psi}_B + (U_1 -\langle U_1 \rangle) \psi_0 \; , \hspace{1cm} U_1 = -\frac{V}{4}
\end{equation}
which satisfies the following relation~\cite{Korobov07,Korobov09}:
\begin{equation}
(E_0 - H_0 ) \tilde{\psi}_B = \left( H'_B - \langle H'_B \rangle \right) \psi_0 \; , \hspace{1cm} H'_B = -(E_0 - H_0)U_1 -U_1 (E_0 - H_0) + H_B.
\end{equation}
One thus obtains
\begin{equation}
\Delta E_b^{(7+)} = 2 \left\langle \tilde{\psi}_B | U_{vp} | \psi_0 \right\rangle + \frac{1}{2} \langle V \rangle \left\langle \psi_0 | U_{vp} | \psi_0 \right\rangle. \label{E7b}
\end{equation}
Finally, the subtracted term is added to the contribution $\Delta E_c$ which is thus redefined as
\begin{equation}
\Delta E_c^{(7+)} = \frac{1}{8} \left\langle \psi_0 | \Delta U_{vp} | \psi_0 \right\rangle - \frac{1}{2} \left\langle \psi_0 | V U_{vp} |\psi_0 \right\rangle
\end{equation}
Integration by parts and use of the Schr\"odinger equation $\Delta \psi_0 = 2 (V - E_0) \psi_0$ provides the following relationship, in which the $\alpha(Z\alpha)^5$-order term has been explicitly canceled out:
\begin{equation}
\Delta E_c^{(7+)} = \frac{1}{4} \left\langle \psi_0 | \mathbf{p} U_{vp} \mathbf{p} | \psi_0 \right\rangle - \frac{E_0}{2} \left\langle \psi_0 | U_{vp} | \psi_0 \right\rangle. \label{E7c}
\end{equation}
The final result is
\begin{equation}
\Delta E_{U}^{(7+)} = \Delta E_a^{(7+)} + \Delta E_b^{(7+)} + \Delta E_c^{(7+)}
\end{equation}
and may be put in the form (see Ref.~\cite{CODATA10} Eq.(46))
\begin{equation}
\Delta E_{U}^{(7+)} =
   \frac{\alpha^5}{\pi} \left[ V_{61} \ln(\alpha^{-2}) + G_{\rm VP}^{(1)}(R) \right] \langle V_{\delta} \rangle
\end{equation}
with $V_{61} = -2/15$. The logarithmic term comes from the logarithmic singularity in $\tilde{\psi}_B$, and should thus be subtracted from $\Delta E_b^{(7)}$:
\begin{equation}\label{vp78}
G_{\rm VP}^{(1)} (R) =
   \pi \Delta E_a^{(7+)} / \langle V_{\delta} \rangle
   + \left[\pi \Delta E_b^{(7+)} / \langle V_{\delta} \rangle - V_{61} \ln(\alpha^{-2})\right]
   + \pi \Delta E_c^{(7+)} / \langle V_{\delta} \rangle.
\end{equation}
Since the initial NRQED approximation is valid up to and including $m\alpha^8$ order, the result of Eq.~(\ref{vp78}) should be accurate to $\mathcal{O}(\alpha^2)$.

\section{Results and conclusion}

We calculated all operator mean values appearing in Eqs.~(\ref{E7a}), (\ref{E7b}) and (\ref{E7c}) for the ground ($1s\sigma$) electronic state of the two-center problem, both for $Z_1 = Z_2 = 1$ for application to H$_2^+$ and HD$^+$, and $Z_1 = 2$, $Z_2 = -1$ for application to antiprotonic helium. The numerical approach has been described previously (see e.g.~\cite{Korobov13}). The following expansion for the $\sigma$ electronic wavefunction is used:
\begin{equation}\label{exp}
   \Psi_0(\mathbf{r}) = \sum^{\infty}_{i=1} C_{i} e^{-\alpha_{i} r_1 - \beta_{i} r_2},
\end{equation}
For $Z_1=Z_2$ the variational wavefunction should be symmetrized
\begin{equation}\label{expsym}
   \Psi_0(\mathbf{r_{1},r_{2}}) = \sum^{\infty}_{i=1}  C_{i}(e^{-\alpha_{i} r_{1} - \beta_{i} r_{2}}\pm  e^{-\beta_{i} r_{1} - \alpha_{i} r_{2}} ),
\end{equation}
where $(+)$ is used to get a {\em gerade} electronic state and $(-)$ is for an {\em ungerade} state, respectively. Parameters $\alpha_{i}$ and $\beta_{i}$ are generated in a quasi-random manner.

The matrix elements of the Uehling potential in such an exponential basis set are not known in analytical form, in contradistinction with the case of the three-body problem~\cite{Karr13}. We thus resorted to numerical integration for all the terms involving $U_{vp}$. To that end we used the approximate form of the Uehling potential presented in~\cite{Fullerton76} which is accurate to at least nine digits.

\begin{figure}[t]
\begin{center}
\begin{tabular}{c@{\hspace{10mm}}c}
\includegraphics[width=0.45\textwidth]{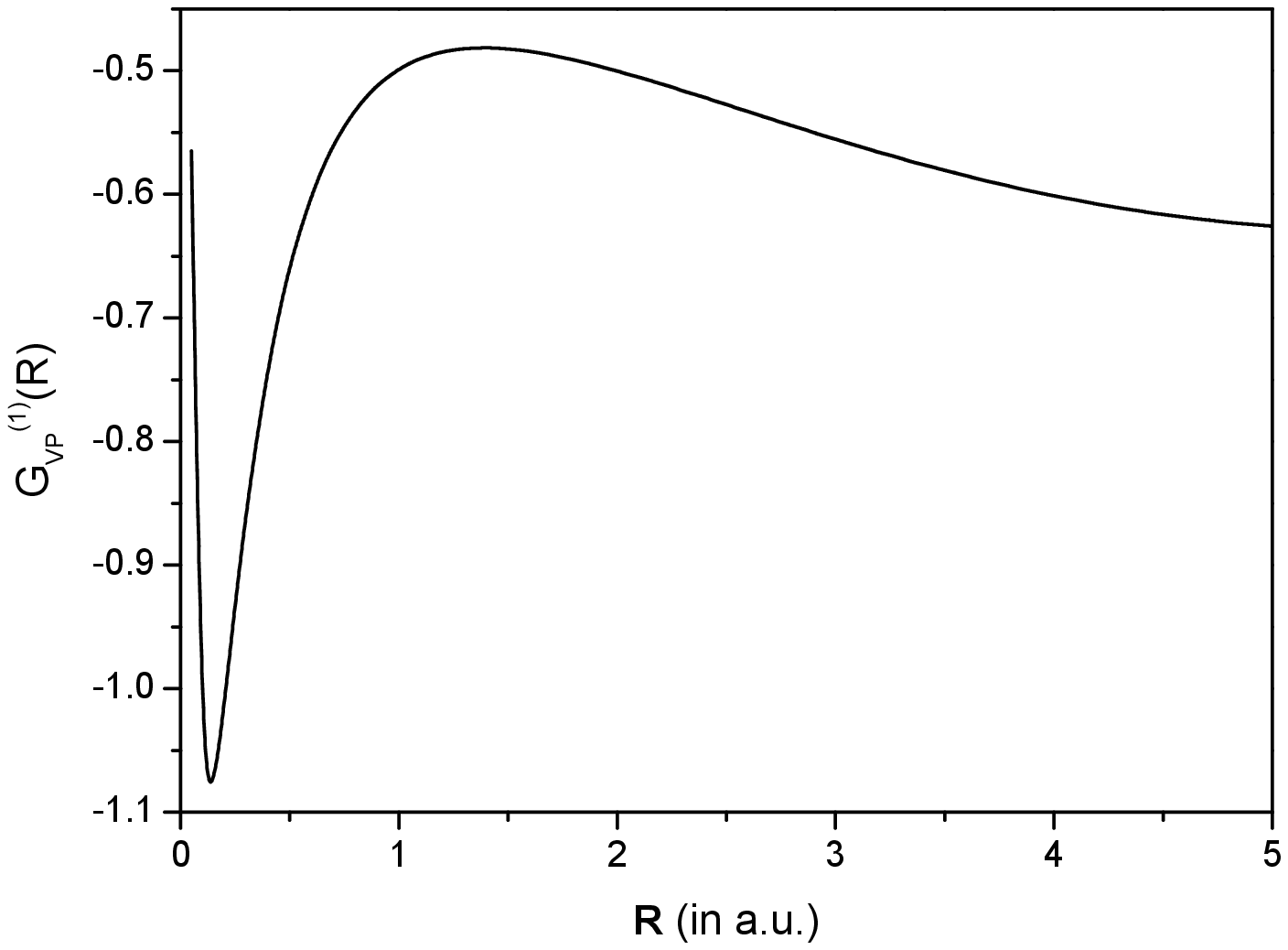}
&
\includegraphics[width=0.45\textwidth]{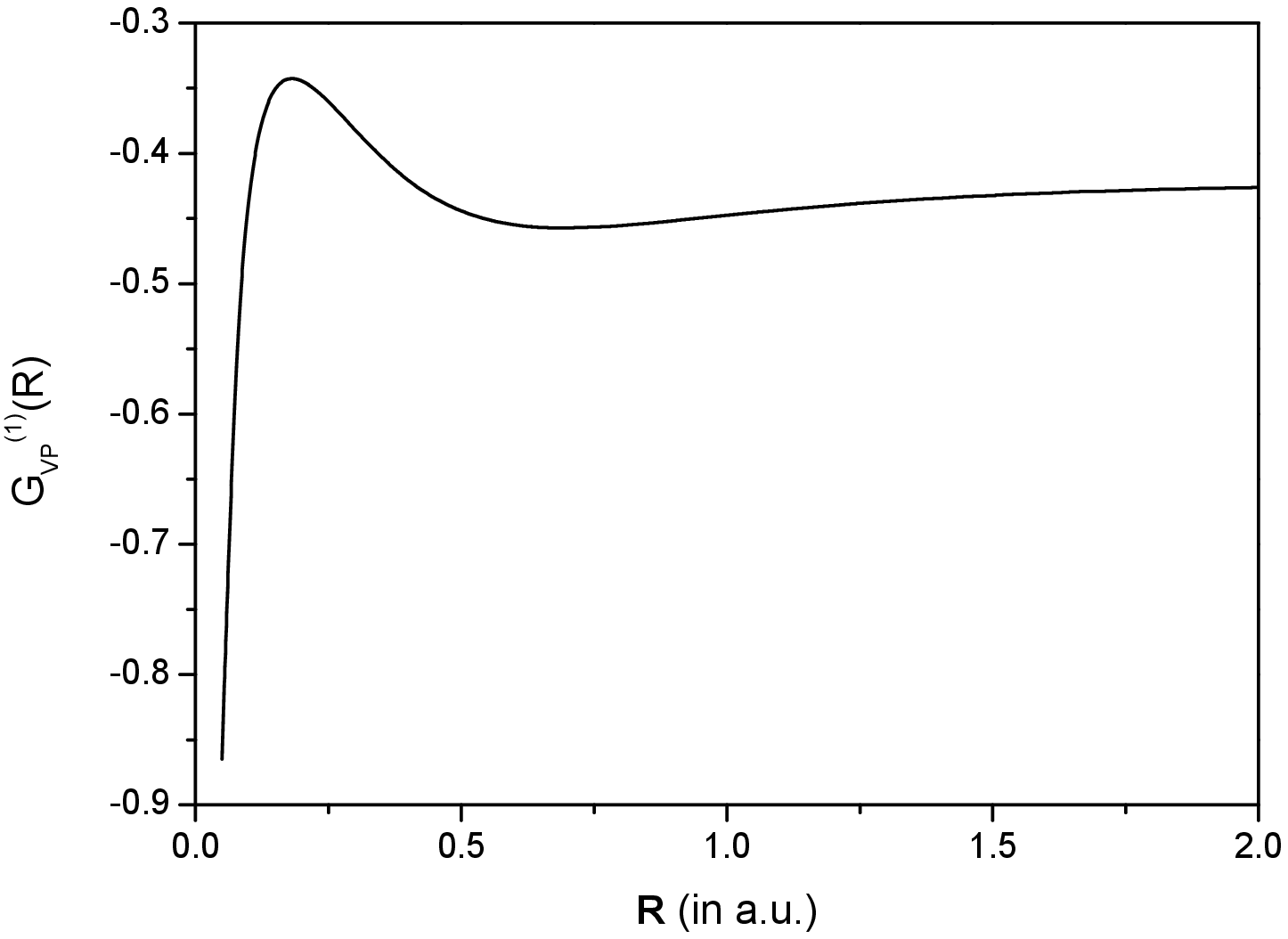}
\end{tabular}
\end{center}
\caption{''Effective potentials'' $G_{\rm VP}^{(1)}(R)$ for the hydrogen molecular ions, $Z_1=Z_2=1$ (left), and antiprotonic helium, $Z_1=2$, $Z_2=-1$ (right). \label{VP_pot}}
\end{figure}

\begin{table}[t]
\begin{center}
\begin{tabular}{@{\hspace{1mm}}c@{\hspace{10mm}}r@{\hspace{5mm}}r@{\hspace{12mm}}r@{\hspace{5mm}}r@{\hspace{4mm}}}
\hline\hline
  & \multicolumn{2}{c}{H$_2^+$\hspace*{8mm}} & \multicolumn{2}{c}{HD$^+$~~} \\
  & this work\!\!\! & LCAO \cite{Korobov14PRA}\!\!\!\!\! & this work\!\!\! & LCAO \cite{Korobov14PRA}\!\!\!\!\! \\
\hline
ground state (kHz) & $-$28.35 & $-$34.73 & $-$28.38 & $-$34.93\\
\hline
 transition (kHz) & 0.42 & 0.94 & 0.37 & 0.82 \\
\hline\hline
\end{tabular}
\end{center}
\caption{Results of numerical calculations of the $G_{\rm VP}^{(1)}$ contribution for the ground states of H$_2^+$ and HD$^+$ and the fundamental transitions: $(v=0,L=0)\to(v'=1,L'=0)$. Comparison is presented with previous estimates made in~\cite{Korobov14PRA} within the LCAO approximation.\label{VP_tab}}
\end{table}

Results are shown in Fig.~\ref{VP_pot}. As can be seen, the values of $G_{\rm VP}^{(1)}(R)$ at $R\to0$ tend to infinity and do not obey the continuity relationship that could be expected, $G_{\rm VP}^{(1)}(R)\to G_{\rm VP}^{(1)}(\mbox{H}_Z\mbox{(1S)})$ where H$_Z$(1S) denotes the 1S state of a hydrogenic atom with nuclear charge $Z = Z_1 + Z_2$. The reason for such a behavior is that the coefficients of the $Z\alpha$ expansion have no physical meaning, and only the sum over all orders matters. Only the complete Uehling potential contribution indeed is a continuous function of $R$ at the united atom limit. The same observation is also valid for the one-loop self energy contribution~\cite{Korobov13}, as well as for higher-order diagrams.

On the contrary, continuity is observed at the other limit, $R\to\infty$. We checked this by direct numerical evaluation of the expressions~(\ref{E7a}), (\ref{E7b}) and (\ref{E7c}) with 1S hydrogenic wavefunctions. The values of $G_{\rm VP}^{(1)}(R)$ at large $R$ converge towards $G_{\rm VP}^{(1)}(\mbox{H}_{Z=1}\mbox{(1S)}) = -0.61845$ in the hydrogen molecular ion case, and towards $G_{\rm VP}^{(1)}(\mbox{H}_{Z=2}\mbox{(1S)}) = -0.42194$ in the antiprotonic helium case.

The last step is numerical integration of the vacuum polarization ''effective'' potentials of Fig.~\ref{VP_pot} over vibrational or heavy particle degrees of freedom to get the energy corrections for individual states. Numerical results for the ground states of H$_2^+$ and HD$^+$ and for the fundamental transitions: $(v=0,L=0)\to(v'=1,L'=0)$, are collected in Table~\ref{VP_tab}. Comparison with the LCAO approximation demonstrates that in case of individual states it may give some reasonable estimate. However, for the transition frequency, due to the slope of the ''effective'' potential at the equilibrium position at $R=2.0$, the difference in contributions from the two states becomes substantially sensitive and the LCAO estimate only gives an order of magnitude. This tendency is less marked in the case of antiprotonic helium, e.g. for the two-photon $(33,32)\to(31,30)$ transition in $^4$He$\bar{p}$ we obtain a shift of 121 kHz, while the LCAO estimate is 98 kHz. That may be explained as follows: the dominating contribution comes from the $1S$ state wave function of hydrogenlike helium ($Z=2$), and the contribution from the antiproton is negligible. However, it is worth noting that for the antiprotonic helium, nonadiabatic effects become essential at this level, and complete three-body calculations are needed to get improved accuracy.

In conclusion, we have calculated the Uehling corrections at orders $m\alpha^7$ and $m\alpha^8$ for the two-center problem. Together with improved numerical calculations of the relativistic Bethe logarithm~\cite{Korobov13}, these results will allow for further improvement of the theoretical accuracy on transition frequencies in H$_2^+$, HD$^+$ and antiprotonic helium.

\section{Acknowledgements}

V.I.K. acknowledges support of the Russian Foundation for Basic Research under Grant No.~12-02-00417-a. This work was supported by \'Ecole Normale Sup\'erieure, which is gratefully acknowledged. J.-Ph. Karr is a member of the Institut Universitaire de France.

\end{document}